\def\year{2020}\relax
\documentclass[letterpaper]{article} 
\usepackage{aaai20}  
\usepackage{times}  
\usepackage{helvet} 
\usepackage{courier}  
\usepackage[hyphens]{url}  
\usepackage{graphicx} 
\usepackage{graphicx}  
\title{Software Testing for Machine Learning}
\author{Dusica Marijan, Arnaud Gotlieb\\ 
Simula Research Laboratory, Norway\\ 
dusica@simula.no, arnaud@simula.no 
}

\providecommand{\keywords}[1]{\textbf{\textit{Index terms---}} #1}

\begin{document}

\maketitle

\begin{abstract}
Machine learning has become prevalent across a wide variety of applications. Unfortunately, machine learning has also shown to be susceptible to deception, leading to errors, and even fatal failures. This circumstance calls into question the widespread use of machine learning, especially in safety-critical applications, unless we are able to assure its correctness and trustworthiness properties. Software verification and testing are established technique for assuring such properties, for example by detecting errors. However, software testing challenges for machine learning are vast and profuse - yet critical to address. This summary talk discusses the current state-of-the-art of software testing for machine learning. More specifically, it discusses six key challenge areas for software testing of machine learning systems, examines current approaches to these challenges and highlights their limitations. The paper provides a research agenda with elaborated directions for making progress toward advancing the state-of-the-art on testing of machine learning.
\end{abstract}


\section{Introduction}
Applications of machine learning (ML) technology have become vital in many innovative domains. At the same time, the vulnerability of ML has become evident, sometimes leading to catastrophic failures\footnote{Tesla failure, www.theguardian.com/technology/2016/jul/01/tesla-driver-killed-autopilot-self-driving-car}. This entails that comprehensive testing of ML needs to be performed, to ensure the correctness and trustworthiness of ML-enabled systems. 

Software testing of ML systems is susceptible to a number of challenges compared to testing of traditional software systems. In this paper, by \textit{\textbf{traditional systems}} we mean software systems not integrating ML, and by \textit{\textbf{ML systems}} we mean software systems containing ML-trained components (e.g self-driving cars, autonomous ships, or space exploration robots). As an example, one such challenge of testing ML systems stems from non-determinism intrinsic to ML. Traditional systems are typically pre-programmed and execute a set of rules, while ML systems reason in a probabilistic manner and exhibit non-deterministic behavior. This means that for constant test inputs and preconditions, an ML-trained software component can produce different outputs in consecutive runs. Researchers have tried using testing techniques from traditional software development \cite{Hutchison2018}, to deal with some of these challenges. However, it has been observed that traditional testing approaches in general fail to adequately address fundamental challenges of testing ML \cite{Helle2016}, and that these traditional approaches require adaptation to the new context of ML. The better we understand current research challenges of testing ML, the more successful we can be in developing novel techniques that effectively address these challenges and advance this scientific field. 

In this paper, we: i) identify and discuss the most challenging areas in software testing for ML, ii) synthesize the most promising approaches to these challenges, iii) spotlight their limitations, and iv) make recommendations of further research efforts on software testing of ML. We note that the aim of the paper is not to exhaustively list all published work, but distill the most representative work. 

\section{Testing ML}
As ML technologies become more pervasive enabling autonomous system functionality, it is more and more important to assure the quality of autonomous reasoning supported by ML. Testing is such a quality assurance activity that aims (in a broad sense) to determine the correctness of the system-under-test, for example, by checking whether the system responds correctly to inputs, and to identify faults which may lead to failures. 

\noindent \textbf{Interpreting "Testing ML"}: Two distinct communities have been studying the concept of \textit{testing ML}, the ML scientific community (MLC) and the software testing community (STC). However, as the two communities study ML algorithms from different perspectives, they interpret the term \textit{\textbf{testing ML}} differently, and we think it is worth noting the distinction. In MLC, testing an ML model is performed to estimate its prediction accuracy and improve its prediction performance. Testing happens during model creation, using validation and test datasets, to evaluate the model fit on the training dataset. In STC, testing an ML system has a more general scope aiming to evaluate the system behavior for a range of quality attributes. For example, in case of integration or system level testing, an ML component is tested in interaction with other system components for functional and non-functional requirements, such as correctness, robustness, reliability, or efficiency. 

\subsection{Challenges of Testing ML}
Challenges of testing ML stem from the innate complexity of the underlying stochastic reasoning. Unlike traditional systems, for which the code is built deductively, ML systems are generated inductively. The logic defining system behavior is inferred from training data. Consequently, a fault could originate not only from a faulty software code, but also errors in training data. However, existing approaches often assume that high quality datasets are warranted, without applying systematic quality evaluation.  
Furthermore, ML systems require advanced reasoning and learning capabilities that can give answers in conditions where the correct answers are previously unknown \cite{Murphy2007}. Even though this may be the case for traditional systems, ML systems have inherent non-determinism which makes them constantly change behavior as more data becomes available, unlike traditional systems \cite{8718214}. Furthermore, for a system containing multiple ML models, the models will affect each other's training and tuning, potentially causing non-monotonic error propagation \cite{Amershi}. 

We elaborate further challenges of testing ML in the following sections. Specifically, we identify six key challenge areas and discuss their implications. We synthesize existing work pertaining to these challenges and provide its structured presentation corresponding to the identified challenges. 

\section{Missing Test Oracles}
Unlike traditional systems which operate pre-programmed deterministic instructions, ML systems operate based on stochastic reasoning. Such stochastic or probability-based reasoning introduces uncertainty in the system response, which gives rise to non-deterministic behavior, including unpredictable or underspecified behavior. Due to non-determinism, ML systems can change behavior as they learn over time. The implications for testing are that system outputs can change over time for the same test inputs. This fact largely complicates test case specification. 

Test cases are typically specified with specific inputs to the system under test and expected outputs for these inputs, known as \textit{\textbf{test oracles}}. However, due to stochastic reasoning, the output of an ML system cannot be specified in advance, rather it is learned and predicted by an ML model. This means that ML systems do not have defined expected values against which actual values can be compared in testing. Thus, the correctness of the output in testing ML cannot be easily determined. While this problem has been known for traditional systems, called "non-testable" systems \cite{Weyuker1982}, ML systems have non-determinism as part of their design, making the oracle problem even more challenging.

An approach that has been considered for non-testable systems are \textit{\textbf{pseudo-oracles}} \cite{Weyuker1982}.  
Pseudo-oracles are a differential testing technique that consists in running multiple systems satisfying the same specification as the original system under test, then feeding the same inputs to these systems and observing their outputs. Discrepancies in outputs are considered indicative of errors in the system under test. A limitation of differential testing is that it can be resource-inefficient as it requires multiple runs of the system, and error-prone, as the same errors are possible in multiple implementations of the system under test \cite{Knight1986}. 

\subsection{Metamorphic Testing}
Metamorphic testing is another approach to testing of software without test oracles. In this approach, a transformation function is used to modify the existing test case input, and produce a new output. If the actual output for the modified input differs from the expected output, it is indicative of errors in the software under test. Metamorphic testing has been applied to machine learning classifiers \cite{Xie2011} \cite{Dwarakanath2018}. However, in testing ML systems with a large input space, writing metamorphic transformations is laborious, and there is a great potential for ML to circumvent this difficulty by automating the creation of metamorphic relationships.

\subsection{Test Data Prioritization}
Since automated oracles are typically not available for testing of big and realistic ML models, there is a great effort involved in manual labeling of test data for ML models. DeepGini \cite{Shi2019} is an initial work on reducing the effort in labeling test data for DNNs by prioritizing tests that are likely to cause misclassifications. The assumption made by DeepGini is that a test is likely to be misclassified if a DNN outputs similar probabilities for each class. The limitation of this approach is that it requires running all tests first, to obtain the output vectors used to calculate the likelihood of misclassification.

\section{Infeasibility of Complete Testing}
ML systems are commonly deployed in application areas dealing with a large amount of data. This creates large and diverse test input space. Unfortunately, testing is rarely able to cover all valid inputs and their combinations to examine the correctness of a system-under-test, and therefore \textit{\textbf{coverage metrics}} are typically applied to select an adequate set of inputs from a large input space, to generate tests, or to assess the completeness of a test set and improve its quality \cite{Marijan2019,Marijan18}. 

\subsection{Test Coverage}
The first attempts to define coverage metrics for testing of neural networks are inspired by the traditional code coverage metrics. A metric called \textit{\textbf{neuron coverage}} was proposed in DeepXplore \cite{Pei2017} for testing deep neural networks (DNN). DeepXplore measures the amount of unique neurons activated by a set of inputs out of the total number of neurons in the DNN. The limitation of this coverage metric is that a test suite that has full neuron coverage (all neurons activated) can still miss to detect erroneous behavior if there was an error in all other DNNs that were part of a differential comparing \cite{Pei2017} used by neuron coverage (DeepXplore leverages the concept of differential testing). Furthermore, it has been shown that neuron coverage can be too coarse a coverage metric, meaning that a test suite that achieves full neuron coverage can be easily found, but the network can still be vulnerable to trivial adversarial examples \cite{Sun2018a}. Sun et al. therefore proposed DeepCover, a testing methodology for DNNs with four test criteria, inspired by the modified condition/decision coverage (MC/DC) for traditional software. Their approach includes a test case generation algorithm that perturbs a given test case using linear programming with a goal to encode the test requirement and a fragment of the DNN. The same author also developed a test case generation algorithm based on symbolic approach and the gradient-based heuristic \cite{Sun2018b}. The difference between their coverage approach, based on MC/DC criterion, and neuron coverage is that the latter only considers individual activations of neurons, while the former considers causal relations between features at consecutive layers of the neural network. 

Neuron coverage has been further extended in DeepGauge \cite{Ma2018}, which aims to test DNN by combining the coverage of key function regions as well as corner case regions of DNN, represented by neuron boundary coverage. Neuron boundary coverage measures how well the test datasets cover upper and lower boundary values. DeepRoad \cite{Zhang2018} is another test generation approach for DNN-based autonomous driving. DeepRoad is based on generative adversarial networks and it generates realistic driving scenes with various weather conditions. DeepCruiser is an initial work towards testing recurrent-neural-network (RNN)-based stateful deep learning \cite{Du2018}. DeepCruiser represents RNN as an abstract state transition system and defines a set of test coverage criteria for generating test cases for stateful deep learning systems. Other approaches were proposed extending the notion of neuron coverage, such as DeepTest \cite{Tian2018} for testing other types of neural networks. DeepTest applies image transformations such as contrast, scaling, blurring to generate synthetic test images. However, such generated images were found to be insufficiently realistic for testing real-world systems.

In summary, a common limitation of techniques based on neuron coverage is that they can easily lead to combinatorial explosion. Ma et al. initiated the work on the adaptation of combinatorial testing techniques for the systematic sampling of a large space of neuron interactions at different layers of DNN \cite{Ma2018a}. This approach can be promising for taming combinatorial explosion in testing of DNN based systems, given that its current limitations are overcome. First, only 2-way interactions of input parameters are supported, while real systems typically have much higher interaction levels of inputs. Second, the approach has been found to face scalability problems for large and complex DNNs.

\subsection{Fuzzing}
Since the input space of DNNs is typically large and highly-dimensional, selecting test data for DNNs can be highly laborious. One approach to deal with this challenge is \textit{\textbf{fuzzing}}, which generates large amounts of random input data that is checked for failures. TensorFuzz is an initial work that applies fuzzing to testing of TensorFlow DNNs \cite{Odena2018}. TensorFuzz uses a coverage metric consisting of user-specified constraints to randomly mutate inputs. The coverage is measured by a fast approximate nearest neighbour algorithm. TensorFuzz has showed to outperform random testing. Another similar approach is DeepHunter \cite{Xie2018}. This is an initial work on automated feedback-guided fuzz testing for DNNs. DeepHunter runs metamorphic mutation to generate new semantically preserved tests, and uses multiple coverage criteria as a feedback to guide test generation from different perspectives. The limitation of this approach is that it uses only a single coverage criteria at the time, not supporting multi-criteria test generation. Moreover, the general limitation of fuzzing is that it cannot ensure that certain test objectives will be satisfied. 

\subsection{Concolic Testing}
To provide more effective input selection that increases test coverage, a concolic testing approach has been proposed in DeepConcolic \cite{Sun2018}. 
The approach is parameterised with a set of coverage requirements. The requirements are used to incrementally generate a set of test inputs with a goal to improve the coverage of requirements by alternating between concrete execution (testing on particular inputs) and symbolic execution. For an unsatisfied requirement, a test input within the existing test suite that is close to satisfying that requirement is identified, based on concrete execution. Later, a new test input that satisfies the requirement is generated through symbolic execution and added to the test suite, improving test coverage. 

\section{Quality of Test Datasets for ML Models}
When training ML models, the quality of the training dataset is important for achieving good performance of the learned model. The performance is evaluated using a test dataset. 

\subsection{Mutation Testing}
To evaluate the quality of test dataset for DNNs, DeepMutation \cite{Ma2018b} proposes an initial work, inspired by traditional mutation testing concepts. DeepMutation first designs a set of mutation operators to inject faults into training data. Then, it retrains models with the mutated training data to generate mutated models, which means that faults are injected in the models. After that, mutated models are tested using a test dataset. Finally, the quality of the test dataset is evaluated by analysing to what extent the injected faults are detected. The limitation of this approach is that it employs basic mutation operators covering limited aspects of deep learning systems, so that the injected faults may not be representative enough of real faults. MuNN \cite{Shen2018} is another mutation testing approach for neural networks, which needs further work for the application on DNNs. Specifically, the authors of the approach showed that neural networks of different depth require different mutation operators. They also showed the importance of developing domain-dependent mutation operators rather than using common mutation operators.

\section{Vulnerability to Adversaries}
ML classifiers are known to be vulnerable to attacks where small modifications are added to input data, causing misclassification and leading to failures of ML systems \cite{Szegedy2014}. Modifications made to input data, called \textbf{\textit{adversarial examples}}, are small perturbations designed to be very close to the original data, yet able to cause misclassifications and to compromise the integrity (e.g. accuracy) of clasiffier. Such attacks have been observed for image recognition \cite{Xie2017}, text \cite{Sato2018}, and speech recognition tasks \cite{Carlini2016} \cite{Carlini2018} \cite{Jia2017}. In the latter, it was shown that adversarially inserted sentences in the Stanford Question Answering Dataset can decrease reading comprehension of ML from 75\% to 36\% of F-measure (harmonic average of the precision and recall of a test).

\subsection{Generating Adversarial Examples}
Adversarial examples can be generated for the purpose of attack or defense of an ML classifier. The former often use heuristic algorithms to find adversarial examples that are very close to correctly classified examples. The latter aim to improve the robustness of ML classifiers. 

Some approaches to adversarial example generation include Fast Gradient Sign Method (FGSM) \cite{Goodfellow2015}, which showed that linear behavior in high-dimensional spaces is sufficient to cause adversarial examples. Later, FGSM was shown to be less effective for black-box attacks \cite{Tramr2017}, and the authors developed RAND-FGSM method which adds random perturbations to modify adversarial perturbations. DeepFool \cite{Moosavi2016} is another approach that generates adversarial examples based on an iterative linearization of the classifier to generate minimal perturbations that are sufficient to change classification labels. The limitation of this approach lies in the fact that it is a greedy heuristic, which cannot guarantee to find optimal adversarial examples. Further, a two-player turn-based stochastic game approach was developed for generating adversarial examples \cite{Wicker2018}. The first player tries to minimise the distance to an adversarial example by manipulating the features, and the second player can be cooperative, adversarial, or random. The approach has shown to converge to the optimal strategy, which represents a globally minimal adversarial image. The limitation of this approach is long runtime. Extending the idea of DeepFool, a universal adversarial attack approach was developed \cite{Moosavi2017}. This approach generates universal perturbations using a smaller set of input data, and uses DeepFool to obtain a minimal sample perturbation of input data, which is later modified into a final perturbation. 

Adversarial examples can be generated with generative adversarial networks, such as AdvGAN \cite{Xiao2018}. This approach aims to generate perturbations for any instance, which can speed up adversarial training. The limitation of the approach is that the resulting adversarial examples are based on small norm-bounded perturbations. This challenge is further addressed in \cite{Song2018} by developing unrestricted adversarial examples. However, their approach exploits classifier vulnerability to covariate shift and is sensitive to different distributions of input data.

\subsection{Countering Adversarial Examples}
To counter adversarial attacks, reactive and proactive defensive methods against adversaries have been proposed. \textit{\textbf{Defensive distillation}} is a proactive approach which aims to reduce the effectiveness of adversarial perturbations against DNNs \cite{Papernot2016}. Defensive distillation extracts additional knowledge about training points as class probability vectors produced by a DNN. The probability vectors are fed back into training, producing DNN-based classifier models that are more robust to perturbations. However, it has been shown that such defensive mechanisms are typically vulnerable to some new attacks \cite{Carlini2017}. Moreover, just like in testing, if a defense cannot find any adversarial examples, it does not mean that such examples do not exist.

\textit{\textbf{Automated verification}} is a reactive defensive approach against adversarial perturbations which analyses the robustness of DNNs to improve their defensive capabilities. Several approaches exist to deal with the robustness challenge. An exhaustive search approach to verifying the correctness of a classification made by a DNN has been proposed \cite{Huang2017}. This approach checks the safety of a DNN by exploring the region around a data point to search for specific adversarial manipulations. The limitation of the approach is limited scalability and poor computational performance induced by state-space-explosion. Reluplex is a constraint-based approach for verifying the properties of DNNs by providing counter-examples \cite{Katz2017}, but is currently limited to small DNNs. An approach that can work with larger DNNs is global optimization based on adaptive nested optimisation \cite{Ruan2018}. However, the approach is limited in the number of input dimensions to be perturbed. A common challenge for verification approaches is their computational complexity. For both approaches \cite{Katz2017} and \cite{Ruan2018}, the complexity is NP-complete. For the former, the complexity depends on the number of hidden neurons, and for the latter, on input dimensions. 

\section{Evaluating the Robustness of ML Models}
To reduce the vulnerability of ML classifiers to adversaries, research efforts are made on systematically studying and evaluating the robustness of ML models, as well as on providing frameworks for benchmarking the robustness of ML models.

\subsection{Robustness Metrics}
Lack of robustness in neural networks raises valid concerns about the safety of systems relying on these networks, especially in safety-critical domains such as transportation, robotics, medicine, or warfare. A typical approach to improve the robustness of a neural network would be to identify adversarial examples that make the network fail, then augment the training dataset with these examples and train another neural network. The robustness of the new network is the ratio between the number of adversarial examples that failed the original network and that were found for the new network \cite{Goodfellow2015}. The limitation of this approach is the lack of objective robustness measure \cite{Bastani2016}. Therefore, a metrics for measuring the robustness of DNNs using linear programming \cite{Bastani2016} was proposed. Other approaches include defining the upper bound on the robustness of classifiers to adversarial perturbations \cite{Fawzi2018}. The upper bound is found to depend on a distinguishability measure between the classes, and can be established independently of the learning algorithms. In their work, Fawzi et al. report two findings: first, non-linear classifiers are more robust to adversarial perturbations than linear classifiers, and second, the depth (rather than breath) of a neural network has a key role for adversarial robustness.

\subsection{Benchmarks for Robustness Evaluation}
There is a difficulty of reproducing some of the methods developed for improving the robustness of neural networks or methods for comparing experimental results, as different sources of adversarial examples in the training process can make adversarial training more or less effective \cite{Goodfellow2016}. To alleviate this challenge, Cleverhans \cite{Goodfellow2016} and Foolbox \cite{Rauber2017} are adversarial example libraries for developing and benchmarking adversarial attacks and defenses, so that different benchmarks can be compared. The limitation of both of these frameworks is that they lack defensive adversarial generation strategies \cite{Yuan2019}. Robust Vision Benchmark \footnote{http://robust.vision/benchmarks/leaderboard} extends the idea of Foolbox, by allowing the development of novel attacks which are used to further strengthen robustness measurements of ML models. Other initiatives include a competition organized at NIPS 2017 conference by Google Brain, where researchers were encouraged to develop new methods for generating adversarial examples and new methods for defense against them \cite{Kurakin2018}.

\subsection{Formal Guarantees over Robustness}
For safety-critical domains which need to comply with safety regulation and certification, it is of critical importance to provide formal guarantees of performance of ML under adversarial input perturbations. Providing such guarantees is a real challenge of most of defense approaches, including the approaches discussed above. Existing attempts in this direction include \cite{Hein2017}, by using regularization in training, and \cite{Sinha2018}, by updating the training objective to satisfy robustness constraints. While these initial approaches are interesting, they can provably achieve only moderate levels of robustness, i.e. provide approximate guarantees. As such, further research advances on providing robustness guarantees for ML models are needed. 

\section{Verifying Ethical Machine Reasoning}
ML systems can be deployed in environments where their actions have ethical implications, for example self-driving cars, and as a consequence, they need to have the capabilities to reason about such implications \cite{Deng2015}. Even more so, if such systems are to become widely socially accepted technologies. While multiple approaches have been proposed for building ethics into ML, the real research challenge lies in building solutions for verifying such machine ethics. This research area has remained largly unaddressed. Existing efforts are limited and include a theoretical framework for ethical decision-making of autonomous systems that can be formally verified \cite{Dennis2016}. The framework assumes that system control is separated from a higher-order decision-making, and uses model checking to verify the rational agent (model checking is the most widely used approach to verifying ethical machine reasoning). However, as a limitation, the proposed approach requires ethics plans that have been correctly annotated with ethical consequences, which cannot be guaranteed. Second, the agent verification is demonstrated to be very slow. For situations where no ethical decision exists, the framework continuous ethical reasoning, negatively affecting overall performance. Third, the approach scales poorly to the number of sensors and sensor values, due to non-deterministic modelling of sensor inputs. Furthermore, the approach cannot provide any guarantees that a rational agent will always operate within certain bounds regardless of the ethics plan. 

Regarding the certification of autonomous reasoning, a proof-of-concept approach \cite{Webster2014} was developed for the generation of certification evidence for autonomous aircraft using formal verification and flight simulation. However, the approach relies on a set of assumptions, such as that the requirements of a system are known, or that they have been accurately translated into a formal specification language, which may not always hold. Finally, ethical machine reasoning should be transparent to allow for checking of the underlying reasoning. These findings emphasize the need for further progress in verifying and certifying ethical machine reasoning.

\section{Summary and Future Directions} 
Software testing of ML faces a range of open research challenges, and further research work focused on addressing these challenges is needed. We envision such further work developing in the following directions. 
\vspace{3pt}\\
\noindent \textbf{Automated test oracles}. Test oracles are often missing in testing ML systems, which makes checking the correctness of their output highly challenging. Metamorphic testing can help address this challenge, and further work is needed on using ML to automate the creation of metamorphic relationships.   
\vspace{3pt}\\
\noindent \textbf{Coverage metrics for ML models}. Existing coverage metrics are inadequate in some contexts. Structural coverage criteria can be misleading, i.e. too coarse for adversarial inputs and too fine for misclassified natural inputs \cite{Li2019}. High neuron coverage does not mean invulnerability to adversarial examples \cite{Sun2018b}. In addition, neuron coverage can lead to input space explosion. Adaptation of combinatorial testing techniques is a promising approach to this challenge, given that progress is made on improving its scalability for real-word ML models. 
\vspace{3pt}\\
\noindent \textbf{Quality of test datasets for ML models}. Evaluation of the quality of datasets for ML models is in its early stages. Adaptation of mutation testing can alleviate this challenge. Common mutation operators are insufficient for mutation testing of DNNs. Instead, domain-specific operators are required.
\vspace{3pt}\\
\noindent \textbf{Cost-effectiveness of adversarial examples}. Generation strategies for adversarial examples need further advancing to reduce computational complexity and improve effectiveness for different classifiers. 
\vspace{3pt}\\ 
\noindent \textbf{Cost-effectiveness of adversarial countermeasures}. Current techniques are mainly vulnerable to advanced attacks. Verification approaches for DNNs to counter adversarial examples are computationally complex (especially constraint-based approaches) and unscalable for real DNNs. More cost-effective verification approaches are required.
\vspace{3pt}\\
\noindent \textbf{Robustness evaluation of ML models}. Metrics for robustness evaluation of ML models and effectiveness evaluation of adversarial attacks need further advancing. Open benchmarks for developing and evaluating new adversarial attacks and defense mechanisms can be useful tools to achieve an improved robustness of defense. Further efforts on understanding the existence of adversarial examples is desired \cite{Yuan2019}.
\vspace{3pt}\\
\noindent \textbf{Certified guarantees over robustness of ML models}. Such guarantees are required for the deployment of ML in safety-critical domains. Current approaches provide only approximate guarantees. Also, further research progress is needed to overcome high computational complexity of producing the guarantees.
\vspace{3pt}\\
\noindent \textbf{Verification of machine ethics}. Formal verification and certification of ethical machine reasoning is uniquely challenging. Further efforts are needed to enable the scalability of these approaches for real systems operating in real-time, and to reach lower computational complexity. In addition, verification approaches may leverage different formal methods, which underlines the open challenge of interoperability between different methods. Finally, research advances on enabling the \textit{\textbf{transparency}} of ethical decision making process is required.
\vspace{5pt}\\
In conclusion, with this paper we hope to provide researchers with useful insights into an unaddressed challenges of testing of ML, along with an agenda for advancing the state-of-the-art in this research area.

\section{ Acknowledgments}
This work is supported by the Research Council of Norway through the project T3AS No 287329.

\fontsize{9.5pt}{10.5pt}
\selectfont 

\bibliographystyle{aaai}
\bibliography{bibliography}

\end{document}